\documentclass[journal]{IEEEtran}
\usepackage{graphicx,amsfonts, amsmath, boldline, amsthm, epstopdf, amssymb, nomencl, adjustbox, stackengine, url, cases,mathtools,mathrsfs, mathrsfs,dsfont, algcompatible,setspace,booktabs,multicol, multirow, color,float, enumitem,cite, cases, mathtools, mathrsfs, ragged2e, adjustbox, stackengine,soul}

\usepackage[dvipsnames]{xcolor}
\usepackage[caption = false]{subfig}
\usepackage[T1]{fontenc}
\usepackage[linesnumbered,lined,boxed,commentsnumbered,ruled,longend]{algorithm2e}

\makenomenclature
\renewcommand\nomgroup[1]{%
  \item[\bfseries
  \ifstrequal{#1}{I}{Sets and Indices}{%
  \ifstrequal{#1}{P}{Parameters}{%
  \ifstrequal{#1}{V}{Variables}{}}}%
]}

\emergencystretch=\maxdimen
\DeclareUnicodeCharacter{2212}{-}

\expandafter\def\expandafter\normalsize\expandafter{%
 	\normalsize
 	\setlength\abovedisplayskip{1.5pt}
 	\setlength\belowdisplayskip{1.5pt}
 	\setlength\abovedisplayshortskip{1.5pt}
 	\setlength\belowdisplayshortskip{1.5pt}
}

\begin{document}






\title{Real-time Optimization of Wind-to-H$_2$ Driven Critical Infrastructures: an Accurate Algorithm based on Feature Space Augmentation}

\title{Online Optimization for Wind-to-H$_2$ Driven Critical Infrastructures: Enhanced Active Constraints Detection with Integer-to-Continuous Approach by Feature Space Augmentation}

\title{\huge Real-time Optimization for Wind-to-H$_2$ Driven Critical Infrastructures: Active Constraints and Integer Variables Prediction Method Enhanced by Feature Space Expansion}

\title{\huge Real-time Optimization for Wind-to-H$_2$ Driven Critical Infrastructures: Active Constraints Detection and Integer-to- Continuous Variables Mapping Enhanced by Feature Space Expansion}


\title{\huge Real-time Optimization for Wind-to-H$_2$ Driven Critical Infrastructures: High-fidelity Active Constraints and Integer Variables Prediction Enhanced by Feature Space Expansion}
 
\author{Mostafa Goodarzi and Qifeng Li,~\IEEEmembership{Senior Member IEEE} \vspace{-0.5cm}

\thanks{This work is supported by U.S. National Science Foundation under Award CBET\#2124849. (Corresponding Author: Qifeng Li)  

The authors are with the Department of Electrical and Computer Engineering, University of Central Florida, Orlando, FL 32816 USA (e-mail: mostafa.goodarzi@knights.ucf.edu, qifeng.li@ucf.edu).
 
}}
    \vspace{-.6cm}

\maketitle

\begin{abstract} 

This paper focuses on developing a real-time optimal operation model for a new engineering system, wind-to-hydrogen-driven low-carbon critical infrastructure (W2H-LCCI), that utilizes wind power to generate hydrogen through electrolysis and combines it with carbon capture to reduce carbon emissions from the power sector. First, a convex mathematical model for W2H-LCCI is proposed, and then optimization models for its real-time decision-making are developed, which are mixed-integer convex programs (MICPs). Furthermore, since this large-scale MICP problem must be solved in real-time, a fast solution method based on active constraint and integer variable prediction (ACIVP) is presented. ACIVP method predicts the binary variable values and the set of limited-number constraints, which most likely contain all of the active constraints, based on historical optimization data. It results in only a small-scale continuous convex optimization problem needing to be solved by optimization solvers for W2H-LCCI real-time optimal operation. To increase the accuracy of the ACIVP method, feature space expansion (FSE) is employed, and a multi-stage ACIVP-FSE method is proposed. The effects of stage design and stage ordering on ACIVP-FSE performance are also discussed. We validate the effectiveness of the developed system and solution method using two water-energy nexus case studies.

\end{abstract}

\begin{IEEEkeywords} Active constraints prediction, Feature expansion, Integer variables prediction, Integrated systems, Low-carbon power generation, Wind to hydrogen 
\end{IEEEkeywords}

\makenomenclature
\mbox{}
    \vspace{-.6cm}
\nomenclature[P$AA$]{$\xi^\mathrm{H}_\mathrm{p}$}{Wind to hydrogen conversion factor}
\nomenclature[P$AAA$]{$\xi^\mathrm{dg}_\mathrm{e}$}{Carbon emission of a diesel generator factor}
\nomenclature[P$AAAA$]{$\xi^\mathrm{we}_\mathrm{p}$}{Energy  for hydrogen production factor}
\nomenclature[P$B$]{$\xi^\mathrm{we}_\mathrm{w}$}{Water for hydrogen production factor}
\nomenclature[P$BB$]{$\xi^\mathrm{fc}_\mathrm{h}$}{Hydrogen for power generation by fuel cell factor}
\nomenclature[P$BB$]{$\xi_\textrm{c}^\textrm{$\chi$}$}{Chemical production by captured carbon factor}
\nomenclature[P$BBB$]{$d_{n}$}{Water demand}
\nomenclature[P$BBBB$]{$\overline{H}$}{Capacity of hydrogen production}
\nomenclature[P$C$]{$p^\mathrm{l},q^\mathrm{l}$}{Active and reactive power load}
\nomenclature[P$CCC$]{$p^\mathrm{rw}$}{Rated power of wind turbine}
\nomenclature[P$DD$]{$r^\mathrm{w}_q$}{Head loss coefficient of pipe $q$}
\nomenclature[P$DDD$]{$r_{ij}, x_{ij}$}{Resistance and reactance of line $ij$}
\nomenclature[P$E$]{$\overline{s}_{ij}$}{Maximum apparent power of line $ij$}
\nomenclature[P$EE$]{$v^\mathrm{ci},v^\mathrm{co}$}{Cut in and cut off wind speed}
\nomenclature[P$F$]{$v^\mathrm{r}$}{Rated wind speed for wind turbine}
\nomenclature[P$G$]{$v_t$}{Wind speed}
\nomenclature[V$A$]{$\chi^\mathrm{S}$}{Chemical Production for selling} 
\nomenclature[V$AA$]{$b^\mathrm{des}, b^\mathrm{p}$}{Binary variables related to desalination and pump}
\nomenclature[V$AAA$]{$b^\mathrm{fc}, b^\mathrm{we}$}{Binary variables related to  FC and water electrolysis}
\nomenclature[V$BB$]{$c^\mathrm{dg}$}{Carbon emission of the diesel generator}
\nomenclature[V$B$]{$c^\mathrm{e}$}{Emitted carbon to the atmosphere}
\nomenclature[V$C$]{$c^\mathrm{s}, c^\mathrm{\chi}$}{Captured carbon for storing and reusing}
\nomenclature[V$CC$]{$f^\mathrm{des}$}{Water production of desalination}
\nomenclature[V$DD$]{$f_n,f_q$}{Water flow of node $n$ and pipe $q$}
\nomenclature[V$DDD$]{$f^\mathrm{wt}$}{Water flow of water tank}
\nomenclature[V$E$]{$h^\mathrm{we}$}{Hydrogen production by water electrolysis}
\nomenclature[V$EE$]{$h^\mathrm{s}$}{Hydrogen value for selling in the hydrogen market}
\nomenclature[V$F$]{$h^\mathrm{ht}$}{Hydrogen value for storing in the hydrogen tank}
\nomenclature[V$FF$]{$I^\textrm{$\chi$}$}{Income from selling the chemical product}
\nomenclature[V$G$]{$I^\mathrm{h}$}{Income from selling hydrogen}
\nomenclature[V$GG$]{$\mathcal{I}_{ij}$}{Square of the current magnitude in line $ij$}
\nomenclature[V$GGG$]{$p_{ij},q_{ij}$}{Active and reactive power of line $ij$}
\nomenclature[V$H$]{$p^\mathrm{des}$}{Power of water desalination}
\nomenclature[V$HH$]{$p^\mathrm{dg},q^\mathrm{dg}$}{Active and reactive power of diesel generator}
\nomenclature[V$HHH$]{$p^\mathrm{fc}$}{Fuel cell output power}
\nomenclature[V$HHHH$]{$p^\mathrm{hs},q^\mathrm{hs}$}{Active and reactive power of hydrogen system}
\nomenclature[V$I$]{$p^\mathrm{sw},q^\mathrm{sw}$}{Surplus active and reactive power of wind farm}
\nomenclature[V$II$]{$p^\mathrm{t}$}{Transferred power to the power network}
\nomenclature[V$III$]{$p^\mathrm{we}$}{Power of water electrolysis}
\nomenclature[V$III$]{$p^\mathrm{wedg}$}{Power of diesel generator for water electrolysis}
\nomenclature[V$JJ$]{$V^\mathrm{ht},V^\mathrm{wt}$}{\hspace{-0.16cm}Volume of hydrogen tank and water tank}
\nomenclature[V$K$]{$\mathcal{V}_{i}$}{Square voltage of bus $i$}
\printnomenclature[1.2cm]


\section{Introduction}
\IEEEPARstart{A}{} promising solution to climate change is the reduction of greenhouse gas emissions from transportation, electricity generation, heating, and industrial sources. Using renewable energy sources (RES), such as wind power, is a widely recognized way to decrease carbon emissions from power generation. However, the integration of wind power into the electrical grid is currently limited by the grid's hosting capacity and may cause negative impacts on grid stability, including voltage fluctuations and flickers \cite{herbert2014review}. Wind-to-hydrogen (W2H) technology converts clean wind energy to hydrogen through water electrolysis in situations where there are limitations and stability issues within the power grid and the need for hydrogen in various applications\cite{schrotenboer2022green,wei2021optimal}. It provides an effective way of increasing wind energy utilization for meeting human energy needs. Recently, researchers investigated integrating W2H with other systems to enhance the overall effectiveness of the integrated system. Some papers examined combining a carbon capture and storage system (CCSS) to reduce carbon emissions.\cite{he2018low,ishaq2020evaluation,ravikumar2020environmental,da2022renewable,yu2022low}. Other studies integrated critical infrastructure systems (CIS) in the W2H system, such as electricity/thermal/gas networks \cite{zhang2020environment}, long-distance freight transportation \cite{sun2021integration}, electric boiler/micro gas turbine/fuel cell (FC) charging station \cite{li2021optimal}, power/gas networks \cite{gu2019power,hu2019stochastic}, and $H_2$ fuel stations \cite{wu2020cooperative}.


In this paper, we discuss integrating W2H with CCSS into CISs, including water, power, and FC generation, to reduce carbon emissions. Since water electrolysis requires significant water, we propose considering the water network as a CIS, unlike the W2H applications mentioned above. The resulting engineering system is called W2H-driven low-carbon critical infrastructures (W2H-LCCI). Besides, since water electrolysis is interconnected with the water distribution system, and water and power systems are linked at the distribution level, our study focuses on the distribution side instead of the generation side, distinguishing it from previous studies. Such W2H application is suitable for remote coastal cities and small islands due to the presence of independent local power systems and significant wind energy resources. For instance, power systems in small islands often rely heavily on diesel generators  \cite{yousefzadeh2020integrated,wijayatunga2016integrating} with high carbon emissions. The $H_2$ produced in the W2H system can be used for fuel cell vehicles (FCVs) which link the transportation with the W2H-LCCI. However, there are not enough FCVs available today to make this link, so our future research will include transportation in the W2H-LCCI as FCVs increase.

Some of the existing studies discussing W2H application did not consider wind uncertainty\cite{wei2021optimal,he2018low,zhang2020environment,sun2021integration}, and instead, they use day-ahead wind speed predictions, which may differ from the real-time measurements \cite{Winddayahead}. We should consider uncertainty when discussing W2H applications because the wind is intermittent. Several studies realize this issue and evaluate wind energy uncertainty based on its intermittent nature \cite{li2021optimal,gu2019power,wu2020cooperative,hu2019stochastic,morteza2023dagging}. Reference \cite{li2021optimal} uses the probabilistic model for wind energy to address the uncertainty related to wind energy production. A robust optimization model, considering wind energy uncertainties, is developed in \cite{gu2019power} to manage power-to-gas technology. Ref \cite{hu2019stochastic} uses the stochastic optimization method to address the uncertainty. The wind uncertainty is modeled in \cite{wu2020cooperative} through a finite number of scenarios generated using the autoregressive moving average model for wind speed. However, these optimization models under uncertainty have some significant limitations. For example, for large engineering systems, their deterministic approximations are generally of extreme-scale and computationally intractable to solve. In stead of making decisions ahead-of-real-time, this paper proposes to solve the optimal operation of W2H-LCCI in real-time which does not rely on long-term (longer than 5 minutes) wind forecast.

Being different from these references, this paper proposes a real-time decision-making (i.e., optimization) scheme to hedge against uncertainty. We develop a convex mathematical model for W2H-LCCI to formulate the W2H-LCCI problem as a real-time decision-making process, which is a large-scale mixed-integer program, which is a large-scale mixed-integer convex program (MICP). To solve such a computationally challenging optimization problem in real-time, we introduce a fast method based on the active constraints and integer variable prediction (ACIVP) \cite{bertsimas2022online}. Based on historical data, the ACIVP-based method maps input parameters into an optimal strategy that includes the optimal values of binary variables and a set of active constraints. The ACIVP enables us to surrogate our MICP problem with a small-scale continuous convex optimization problem that can be rapidly solved by mature solvers, such as Mosek, Gurobi, etc.



Although the ACIVP-based solution process is a quick method, its confidence relies on the accuracy of predicting the active constraints and optimal binary variables. If the accuracy is not high enough, the results obtained by this solution process cannot be trusted. Since considering more features in the learning process can enhance prediction performance, we can improve the accuracy by expanding the feature space\cite{liu2019feature}. After solving offline optimization problems, a considerable amount of raw data becomes available for developing new features. We apply feature space expansion (FSE) and propose a multi-stage ACIVP-FSE method to correct the mapping model learning error in the ACIVP-based method and increase accuracy. Additionally, we discuss the effects of stage design and stage ordering on ACIVP-FSE performance. The main contributions of this paper are:

\begin{itemize}
\item We propose a real-time scheme and develop an optimization model for W2H-LCCIs of coastal cities/small islands, in which CIs are composed of water and power distribution networks.

\item To solve the developed optimization problem which is a large-scale MICP, we introduce the ACVIP-based method to surrogate the initial problem with a smaller-scale continuous convex optimization problem that can be solved quickly.

\item We enhance the confidence of the proposed solution method by improving its accuracy through FSE. An efficient FSE, which is attained by employing a well-designed multi-stage predictive framework, can correct mapping model learning errors and enhance accuracy. Furthermore, we examine the impact of stage design and stage ordering within the multi-stage ACIVP-FSE approach.

\end{itemize}


The rest of this paper is organized as follows. Section \ref{sec: LCEWH_MODEL_DESIGN} presents the system design and mathematical model of the W2H-LCCI for remote coastal cities and small islands. The optimization models for real-time operation are discussed in section \ref{sec: LCEWH_Optmodel}. Section \ref{sec: LCEWH_solmeth} explains the online solution method. Section \ref{sec:casestudy} presents two case studies to validate this method. Finally, conclusions are drawn in section \ref{sec: conclusions}.

\section{Mathematical Models of W2H-LCCI}  \label{sec: LCEWH_MODEL_DESIGN} 
\begin{figure}[!t] 
  \centering
{\includegraphics[width=.475\textwidth]{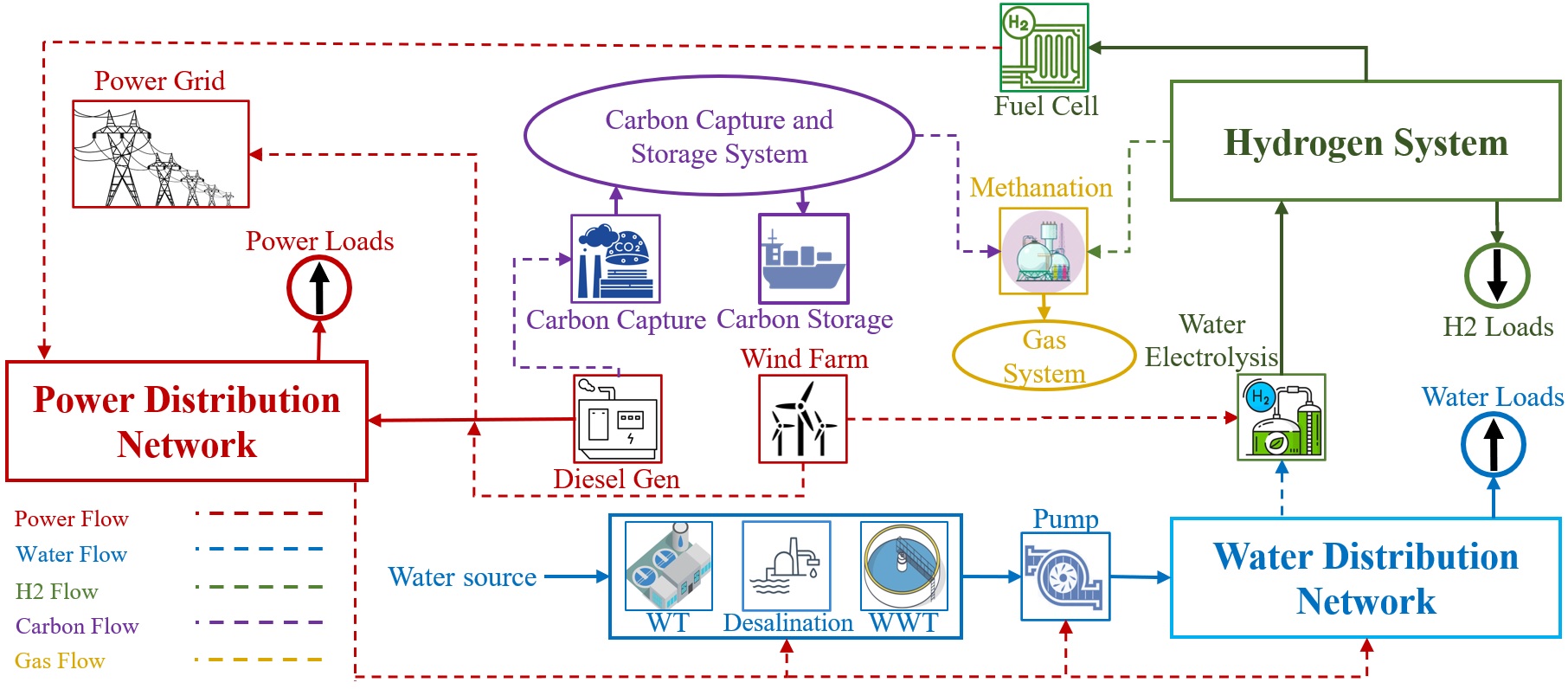}}
 \centering
           \vspace{-.3cm}
    \caption{Wind-to-Hydrogen Driven Low Carbon Critical Infrastructures }
          \vspace{-.5cm}
  \label{pic:bigW2H-LCCI}
\end{figure}

This section explains the different components and the mathematical model of a W2H-LCCI. Fig. \ref{pic:bigW2H-LCCI} shows a typical W2H-LCCI for remote coastal cities/small islands that includes a power section (red), a water section (blue), a hydrogen section (green), a CCSS system (purple), and a methanation system (yellow). A wind farm provides the energy for electrolyzing water, and excess wind energy is fed into the power distribution system for flexible loads like electricity-driven water facilities. When there is a surplus of wind energy, the power grid can serve as a bulk energy storage system. Hydrogen generated by electrolysis is used in various applications, including methanation to reuse captured carbon, converting unpredictable wind energy into controllable energy through FC units, and satisfying the hydrogen network's demand. In the following subsections, we explain the mathematical models of different components of the W2H-LCCI.

\subsection{Power Section} \label{sec:PFPower} 

Power flow in the PDN can be modeled using several formulations. The Distflow model involves bus variables and branch variables and can be used to model both active and reactive power flow in the PDN. The power section of the W2H-LCCI system is expressed with (\ref{eq_PDN1}) to (\ref{eq_netPower}).
\begin{subequations} \label{eq_PDN}
	\begin{align}
    &p^\textrm{wind}_{t}= \begin{cases}
      0, &   v_t \leq v^\textrm{ci} \\
    p^\textrm{rw} ({v_t-v^\textrm{ci} \over v^\textrm{r}-v^\textrm{ci}})^3, &  v^\textrm{ci} \leq  v_t \leq v^\textrm{r}\\
      p^\textrm{rw}, &  v^\textrm{r} \leq  v_t \leq v^\textrm{co}\\
      0, &   v_t \geq v^\textrm{co}. \label{eq_wind_2}
    \end{cases}\\
&\mathcal{V}_{i,t}-\mathcal{V}_{j,t}=2(r_{ij}p_{ij,t}+x_{ij}q_{ij,t})-(r_{ij}^\mathrm{2}+x_{ij}^\mathrm{2}){\mathcal{I}}_{ij,t}, \label{eq_PDN1}\\
&p_{ij,t}^2+q_{ij,t}^2=\mathcal{V}_{i,t}{\mathcal{I}}_{ij,t},\label{eq_PDN2}\\
&p_{ij,t}^2+q_{ij,t}^2\leq {\overline{s}_{ij}}^2,\label{eq_PDN3}\\
&\sum\limits_{k}(p_{ki,t})+r_{ij}{\mathcal{I}}_{ij,t}-p_{ij,t}=p^\mathrm{dg}_{i,t}+p^\mathrm{sw}_{i,t}+p^\mathrm{hs}_{i,t}-p^\mathrm{l}_{i,t},\label{eq_PDN4}\\
&\sum\limits_{k}(q_{ki,t})+x_{ij}{\mathcal{I}}_{ij,t}-q_{ij,t}=q^\mathrm{dg}_{i,t}+q^\mathrm{sw}_{i,t}+q^\mathrm{hs}_{i,t}-q^\mathrm{l}_{i,t},\label{eq_PDN5}\\
&\resizebox{.89\hsize}{!}{$\underline{p}^\mathrm{dg}_{i},\underline{q}^\mathrm{dg}_{i},\underline{\mathcal{V}}_{i},\underline{\mathcal{I}}_{ij} \leq p^\mathrm{dg}_{i,t},q^\mathrm{dg}_{i,t},\mathcal{V}_{i,t},\mathcal{I}_{ij,t}\leq \overline{p}^\mathrm{dg}_{i},\overline{q}^\mathrm{dg}_{i},\overline{\mathcal{V}}_{i},\overline{\mathcal{I}}_{ij}$}\label{eq_PDN6}\\
    &p_t = p^\textrm{wind}_t - (p^\textrm{we}_t + p^\textrm{sw}_t) + p^\textrm{wedg}_t. \label{eq_netPower} 
	\end{align}
\end{subequations}
 The wind farm power can be calculated by (\ref{eq_wind_2}) \cite{xu2020optimized}. Constraints (\ref{eq_PDN1}) to (\ref{eq_PDN3}) are related to Ohm's law. The nodal balance of active and reactive power can be determined by (\ref{eq_PDN4}) and (\ref{eq_PDN5}), respectively. The upper and lower bounds for variables are described by (\ref{eq_PDN6}). The W2H-LCCI may not be able to capture all wind energy due to its volatility and unpredictable characteristics. The power grid can serve as a large energy storage system for capturing excess wind power, as shown with (\ref{eq_netPower}). In this constraint, $p^\textrm{wedg}_t$ represents the power of a diesel generator that is used for water electrolysis when wind power is unavailable, and the technology prevents it from being immediately shut down.

\subsection{Water Section} \label{sec:PFWater}
A convex-hull model for the water section of the W2H-LCCI, including water desalination, mass flow conservation law, pipe flow, water pumps, water tank, and pressure-reducing valves, is given \cite{goodarzi2022evaluate,goodarzi2022security}: 
\begin{subequations} \label{eq3}
	\begin{align}
   &\sum\limits_{m}f_{nm,t} = f^\mathrm{des}_{n,t}-d_{n,t}+f^\mathrm{wt}_{n,t}, \label{eq3_1}\\
   &\widehat{Y}
\begin{cases}
      \leq (2\sqrt{2}-2)r^\mathrm{w}_{q}\overline{f}_{q}f_{q,t}+(3-2\sqrt{2})r^\mathrm{w}_{q}\overline{f}_{q}^2, \\
      \geq (2\sqrt{2}-2)r^\mathrm{w}_{q}\underline{f}_{q}f_{q,t}-(3-2\sqrt{2})r^\mathrm{w}_{q}\underline{f}_{q}^2,  \\
      \geq 2r^\mathrm{w}_{q}\overline{f}_{q}f_{q,t}-r^\mathrm{w}_{q}\overline{f}_{q}^2,  \\
      \leq 2r^\mathrm{w}_{q}\underline{f}_{q}f_{q,t}+r^\mathrm{w}_{q}\underline{f}_{q}^2.
    \end{cases} \label{eq3_2}\\
   &p^\textrm{des}_t = e\times f^\textrm{des}_t ,\label{eq_Water_1}\\
   &e = \begin{cases}
      e_1, &   0\leq f^\mathrm{des}_{t} \leq 0.25 b^\mathrm{des}_{t}\overline{f}^\mathrm{des} \\
      e_2, &   0.25b^\mathrm{des}_{t}\overline{f}^\mathrm{des}\leq f^\mathrm{des}_{t} \leq 0.5b^\mathrm{des}_{t} \overline{f}^\mathrm{des}\\
      e_3, &   0.5b^\mathrm{des}_{t}\overline{f}^\mathrm{des}\leq f^\mathrm{des}_{t} \leq 0.75b^\mathrm{des}_{t}\overline{f}^\mathrm{des}\\
      e_4, &   0.75b^\mathrm{des}_{t}\overline{f}^\mathrm{des}\leq f^\mathrm{des}_{t} \leq b^\mathrm{des}_{t}\overline{f}^\mathrm{des}
    \end{cases}\label{eq_Water_2}\\
     &\widehat{Y}+y^{\mathrm{G}}_{q,t}-r^\mathrm{w}_{q}(f_{q,t})^2 \ge M(b^\mathrm{p}_{q,t}-1),\label{Convwatp1}\\
&\widehat{Y}+y^{\mathrm{G}}_{q,t}-r^\mathrm{w}_{q}\overline{f}_{q}f_{q,t} \le M(1-b^\mathrm{p}_{q,t}),\label{Convwatp2}\\
 &0 \le f_{nm} \le b^\mathrm{p}_{p,t} \overline{f}_{p}.\label{Convwatp3}\\
&V_{n,t+1}^\mathrm{wt}=V_{n,t}^\mathrm{tw}+f_{n,t}^\mathrm{wt}, \label{eq3_6}\\      	
&A_{n}^\mathrm{wt} (y_{n,t+1}^\mathrm{wt} - y_{n,t}^\mathrm{wt}) = f_{n,t}^\mathrm{wt}, \label{eq3_7}\\
&-PR \leq \widehat{Y} \leq PR, \label{eq3_8}\\
&\eta p^\mathrm{p}_{i,t} \ge 2.725\times (a_1(f_{q,t})^2+a_0f_{q,t})\label{ConvPump_1}\\
&\eta p^\mathrm{p}_{i,t} \le 2.725\times (a_1\overline f_{q}+a_0) f_{q,t},\label{ConvPump_2}\\
&\underline{f}^\mathrm{des}_{n}, \underline{f}^\mathrm{wt}_{n},\underline{V}^\mathrm{wt}_{n} \leq f^\mathrm{des}_{n,t}, f^\mathrm{wt}_{n,t},V^\mathrm{wt}_{n,t} \leq \overline{f}^\mathrm{des}_{n}, \overline{f}^\mathrm{wt}_{n},\overline{V}^\mathrm{wt}_{n},\label{eq3_4}\\
&\underline{y}_{n}, \underline{f}_{p} \leq y_{n,t}, f_{p,t},\leq \overline{y}_{n}, \overline{f}_{p},\label{eq3_5}
	\end{align}
\end{subequations}
where $\widehat{Y}= y^c_{n,t}-y^c_{m,t}+h_{q}$, and $q$ is the pipe between node $n$ and node $m$. The equality of water injection and water output at each node is guaranteed by (\ref{eq3_1}). Constraint (\ref{eq3_2}) shows a convex-hull model for head loss along a regular pipe. Equations (\ref{eq_Water_1}) and (\ref{eq_Water_2}) show a model of water desalination. The convex model of a pipe with a pump is expressed by (\ref{Convwatp1}) to (\ref{Convwatp3}). Each tank is modeled as a node using (\ref{eq3_6}) and (\ref{eq3_7}). Pressure-reducing valve is modeled by (\ref{eq3_8}) to control the water head pressure and the convex model of a pump is modeled by (\ref{ConvPump_1}) and (\ref{ConvPump_2}). The upper and lower levels of the variables are shown by (\ref{eq3_4}) and
(\ref{eq3_5}).

\subsection{Carbon Capture and Storage Section}\label{sec:CarbonCSS}
The diesel generator of the proposed system should utilize a CCS to reduce carbon emissions. Following is a model of carbon dioxide emissions resulting from diesel generation, which consists of three parts: emitted parts, ones reused for chemical production, and ones stored.
\begin{subequations} \label{eq_CCS}
	\begin{align}
&c_t^\textrm{dg} = \xi^\textrm{dg}_\textrm{c}p^\textrm{dg}_t,\label{eq_CCS_1}\\
&c^\textrm{e}_t = c_t^\textrm{dg} - c_t^\textrm{s} - c^{\chi}_t,\label{eq_CCS_2}
	\end{align}
\end{subequations}

\subsection{Hydrogen Section}\label{sec:Hydrogen}
The following mathematical formulations describe the hydrogen section of the W2H-LCCI, which includes water electrolysis, an FC unit, and a hydrogen tank:
\begin{subequations} \label{eq_H2}
	\begin{align}
&h^\textrm{we}_t = \xi^\textrm{we}_\textrm{p} p^\textrm{we}_t \label{eq_H2_1}\\
&d^\textrm{we}_t = \xi^\textrm{we}_\textrm{w}  h^\textrm{we}_t,\label{eq_H2_2}\\
&V^\textrm{ht}_{t+1} = (1-\xi^\textrm{dsp})V^\textrm{ht}_{t} + (h^\textrm{we}_t - h^\textrm{fc}_t - h^\textrm{d}_t),\label{eq_H2_3}\\
&p^\textrm{fc}_t = \xi^\textrm{fc}_\textrm{h} h^\textrm{fc}_t,\label{eq_H2_4}\\
&(p^\textrm{we}_t - p^\textrm{fc}_t)^2 + (q^\textrm{hs}_t)^2 \leq (\overline{s}^\textrm{hs})^2,\label{eq_H2_5}\\
&b^\textrm{fc}_t\underline{h}^\textrm{fc}, b^\textrm{we}_t\underline{p}^\textrm{we} \le h^\textrm{fc}_t, {p}^\textrm{we}_t \leq b^\textrm{fc}_t\overline{h}^\textrm{fc}, b^\textrm{we}_t\overline{p}^\textrm{we},\label{eq_H2_6}\\
&\underline{V}^\textrm{ht} \leq  V^\textrm{ht}_t\leq  \overline{V}^\textrm{ht}\label{eq_H2_7}\\
&b^\textrm{we}_t + b^\textrm{fc}_t \leq 1,\label{eq_H2_8}
	\end{align}
	\begin{align}
&\sum_{l=0}^{T^s} \biggl(b^\textrm{we}_{t+l} \biggl)
\begin{cases}
       = 0 , \,\,\,\,\,\, \text{if}\,\, \,\,b^\textrm{we}_{t} - b^\textrm{we}_{t-1} < 0 \\  
      \geq 0 , \,\,\,\,\,\, \text{if}\,\, \,\,b^\textrm{we}_{t} - b^\textrm{we}_{t-1} \geq 0
    \end{cases}, \label{eq:AEL}
	\end{align}
\end{subequations}
where (\ref{eq_H2_1}) shows hydrogen production and (\ref{eq_H2_2}) represents the required water for the electrolysis process. A mass balance equation for $H_2$ is presented by (\ref{eq_H2_3}), which takes into account dissipation rates and the demand for $H_2$ in the HDN. Constraint (\ref{eq_H2_4}) refers to the $H_2$ consumption level of FC units based on efficiency and conversion factors, and (\ref{eq_H2_5}) presents the $H_2$ systems' inverter for reactive power support \cite{haggi2022proactive}. The  upper and lower level of water electrolysis power and hydrogen tank volume, and hydrogen rate of the FC unit are shown in (\ref{eq_H2_6}) and (\ref{eq_H2_7}). The simultaneous operation of electrolysis and FC units is avoided by (\ref{eq_H2_8}). As alkaline electrolysis (AEL) is the most mature technology and has a lower installation cost than other water electrolysis technologies \cite{scolaro2022optimizing}, we chose to incorporate this technology into the W2H-LCCI framework. Since AEL starting up takes approximately 30 to 50 minutes each\cite{gotz2016renewable,chi2018water}, we consider a 1-hour interval sufficient to change the ON/OFF status. The ON/OFF status of the AEL will be 0 for $T^s$ continuous time intervals, totaling 1 hour, after the switch has been changed from ON to OFF, as indicated in (\ref{eq:AEL}).

\subsection{Methanation System}
The methanation system combines the captured carbon with hydrogen to generate $CH_4$ using a Sabatier reaction\cite{gahleitner2013hydrogen}.
Based on the chemical equation for this process ($4H_2 + CO_2 \rightarrow CH_4 + 2H_2O$), we need $182$ $g$ of hydrogen to recycle $1$ $kg$ of carbon, which can be produced from $1.64$ $liters$ of water. The following equations show income from selling chemical production. 
 \begin{align}
    I^\textrm{$\chi$}_t = \rho^\textrm{$\chi$} \, \xi_\textrm{c}^\textrm{$\chi$} \, c^\textrm{$\chi$}_t. \label{eq_Chemical_S}
	\end{align}

\section{Optimization Models for real-time operation} \label{sec: LCEWH_Optmodel}
The proposed W2H-LCCI system requires actual wind speed to address wind uncertainty in real-time optimal operation mode. This paper assumes that 5-minute-ahead wind predictions are accurate enough to be taken as real-time wind speed. Since the FC, desalination plant, and water pump can rapidly change their operational status \cite{pesaran2005pem,goodarzi2022evaluate}, controlling these devices is relatively straightforward. In contrast, AEL technology (which requires startup time and minimum working capacity) makes optimization more challenging. Optimization models for other water electrolysis technologies need to be different, such as solid oxide and proton exchange membrane (PEM) electrolysis. This section presents two optimization models for the real-time operation of the proposed W2H-LCCI according to two different water electrolysis technologies. The objective function of these optimization models that minimizes carbon emissions and wind power transfer to the power grid, while maximizing revenue from the sale of chemical production is shown as:
	\begin{align} \label{eq_Objective} 
    \sum_{t=1}^{T}\,\,\biggl( a_1 p_t + a_2 p^\mathrm{dg}_{t} + a_3 c_t^\textrm{e} + a_4 c_t^\textrm{s} 
 -  I^\textrm{$\chi$}_t \biggl),
	\end{align}
where $a_1$ to $a_4$ are the parameters that show the penalty or cost. Solution methods to achieve real-time optimal solutions at high speed are discussed in Section \ref{sec: LCEWH_solmeth}.


\subsection{Optimization Model for Real-Time Decision-Making of W2H-LCCI with AEL}
\begin{figure}[!t]  
  \centering
\includegraphics[width=0.478\textwidth]{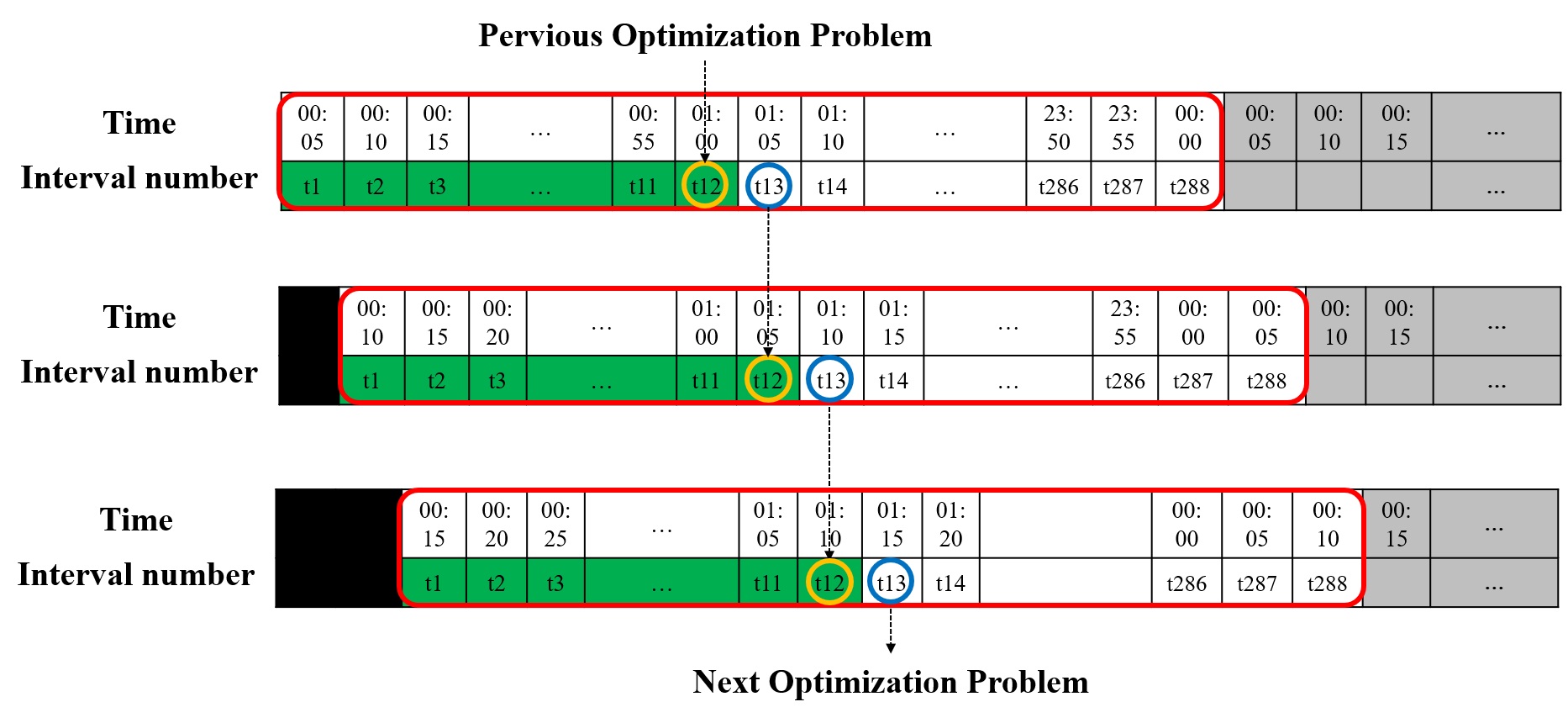}
 \centering
     \vspace{-.2cm}
    \caption{Real-time optimization model for the W2H-LCCI with AEL: the green circles show the ON/OFF status of the AEL which should be fixed within the next $T^\textrm{s}$ time intervals. The blue circles are obtained in the current optimization problem ($\textbf{b}_{T^\textrm{s}+1}^\textrm{\textbf{we*}}$) and will be fixed in the following optimization problem (yellow circles)}
  \label{pic:OSAEL} 
  \vspace{-.5cm}
\end{figure}

AEL electrolysis is the most common water electrolysis technology and has a cost advantage in terms of installation, but it needs one hour for switching. Additionally, AEL electrolysis requires a minimum operating capacity of 20\%. Due to the presence of devices such as water and hydrogen tanks that can store and utilize these resources throughout the day, an operation horizon of 24 hours should be considered in the optimization model of W2H-LCCI. The status of AEL for the next hour, because of the switching time limitation, is determined in every time step. Indeed, in each time step, we know the status of AEL for the following $T^\textrm{s}$ time steps (here, $T^\textrm{s}$ equals 12 because the switching time is 60 minutes and the time step is 5 minutes). The AEL status for the next hour is found and fixed for the next time step. Other binary variables can be changed in the following time step optimization. Algorithm \ref{alg:AEL} and Fig. \ref{pic:OSAEL} illustrate the optimization model of the W2H-LCCI with AEL.
\begin{algorithm}[!b]
\label{alg:AEL}
\captionsetup{font={small,sf,bf}, labelsep=newline}
  \caption{\footnotesize Optimal operation of the W2H-LCCI with AEL.} 
\footnotesize 
\begin{algorithmic}[1]
\justifying
\STATE Input day ahead forecasted $W_0 = \{v^{\textrm{F}_0}_1,...,v^{\textrm{F}_0}_{T}\}$, \\ $P^L_0 = \{p^{\textrm{l,F}_0}_1,...,p^{\textrm{l,F}_0}_{T}\}$ and $d_0 = \{d^{\textrm{F}_0}_1,...,d^{\textrm{F}_0}_{T}\}$
\STATE Solve mixed integer problem and find the optimal solution to find\\ the ON/OFF status for the first hour of the day 
\FOR{i=1:T every t minutes}
\STATE Input real-time and day-ahead wind speed and power demand  \\ $W_i = \{v^{\textrm{F}_i}_i, v^{\textrm{F}_i}_{i+1},...,v^{\textrm{F}_i}_{i+T}\}$ , $v^{\textrm{F}_i}_i = v^{\textrm{R}}_i $, \\ 
$P^\textrm{l}_i = \{p^{\textrm{l,F}_i}_i, p^{\textrm{l,F}_i}_{i+1},...,p^{\textrm{l,F}_i}_{i+T}\}$ , $p^{\textrm{l,F}_i}_i = p^{\textrm{l,R}}_i $, \\ 
$d_i = \{d^{\textrm{F}_i}_i, d^{\textrm{F}_i}_{i+1},...,d^{\textrm{F}_i}_{i+T}\}$ , $d^{\textrm{F}_i}_i = d^{\textrm{R}}_i $.
\STATE Fix binary variables for $T^\textrm{s}$ upcoming times to the values \\ that obtained in the previous optimization problem, $B^t_i = B^{t-1}_i$. 
\STATE Apply \textbf{Algorithm} \textbf{\ref{alg:OCTsOMIO}} mentioned in \textbf{subsection \ref{sec: LCEWH_solmeth}} to find the \\surrogate optimization problem.
\STATE Solve the optimization problem obtained in \textbf{Step 6}
\STATE Update the optimal value for binary variables for the next \\ optimization problem.
\ENDFOR
\end{algorithmic}
\end{algorithm}
The optimization model for W2H-LCCI with AEL is shown as follows:
\begin{align} \label{eq:AELOpMo}
& \textbf{min} \,\,\,\,\ (\ref{eq_Objective}) \nonumber  \\
& s.t \,\,\,\,\,\,\,\ (\ref{eq_PDN}) - (\ref{eq_Chemical_S}) \nonumber  \\
& \,\,\,\,\,\,\,\,\,\,\,\,\,\,\,\,  b_t^\textrm{we} =  \textbf{b}_t^\textrm{\textbf{we*}}, \,\,\,\, \forall \,\, t \in \{1,...,T^\textrm{s}\}, \nonumber\\
& \,\,\,\,\,\,\,\,\,\,\,\,\,\,\,\, b_t^\textrm{p}, b_t^\textrm{we},b_t^\textrm{des},b_t^\textrm{fc} \in \{0,1\}.
	\end{align}
\subsection{Optimization Model for Real-time Decision-Making of Future W2H-LCCI with PEM}
W2H-LCCI may use other water electrolysis technologies in the near future, such as solid oxide and PEM electrolysis. Solid oxide electrolysis is still in its research phase, so this study does not further investigate solid oxide electrolysis. Instead, this paper focuses on PEM water electrolysis which is a relatively new technology with a faster switching time and does not require a minimum load for operation \cite{undertaking2017study,scolaro2022optimizing}. As a result, we only need to solve the following optimization problem at each time step:
\begin{align} \label{eq:PEMOpMo}
& \textbf{min} \,\,\,\,\ (\ref{eq_Objective}) \nonumber  \\
& s.t \,\,\,\,\,\,\,\ (\ref{eq_PDN})-(\ref{eq_Chemical_S}), \nonumber \\
& \,\,\,\,\,\,\,\,\,\,\,\,\,\,\,\,\, b_t^\textrm{p}, b_t^\textrm{we},b_t^\textrm{des},b_t^\textrm{fc} \in \{0,1\}.
	\end{align}




\section{The proposed online solution method}
\label{sec: LCEWH_solmeth}
\subsection{Fast Solution Method Based on ACIVP} \label{sec: SolutionMethod1}
Since problems (\ref{eq:AELOpMo}) and (\ref{eq:PEMOpMo}) must be solved in real-time, the computation efficiency of the solution method is critical. Even though we have convexified the non-convex constraints in these problems, the binary variables and the huge size still make the problem computationally challenging using conventional optimization methods. This section presents the ACIVP-based approach which is a fast solution method to solve the real-time optimization problem for W2H-LCCI operation. ACIVP surrogates (\ref{eq:AELOpMo}) and (\ref{eq:PEMOpMo}) with new optimization problems which contain active constraints and optimal values of binary variables. We define active constraints as the set of constraints that are satisfied as equalities at optimality \cite{bernau1990active}. By identifying the active constraints, all other nonactive constraints can be removed since nonactive constraints do not influence the optimal solution. ACIVP employs data-driven methods to predict active constraints and optimal values of binary variables. Data from solved offline optimization problems are used for this purpose, allowing us to apply MCIP to real-time operations that were previously impractical. ACIVP method surrogates the original problems by forecasting binary variables ($b_t^\textrm{p} =  \textbf{b}_t^\textrm{\textbf{p*}},b_t^\textrm{we} = \textbf{b}_t^\textrm{\textbf{we*}},
b_t^\textrm{des} = \textbf{b}_t^\textrm{\textbf{des*}},
b_t^\textrm{fc} = \textbf{b}_t^\textrm{\textbf{fc*}}$) and active constraints using data-driven approaches. 
\begin{figure}[!t]  
  \centering
\includegraphics[width=0.478\textwidth]{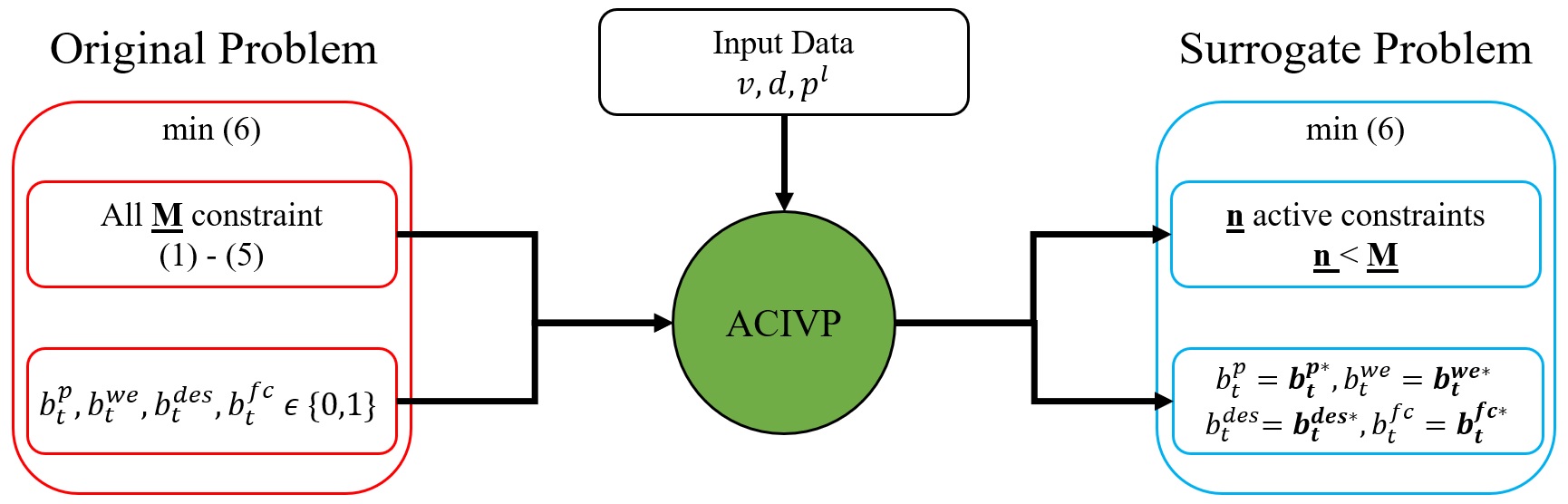}
 \centering
     \vspace{-.2cm}
    \caption{Surrogate the original problem using ACIVP-based method: input data are wind, water demand, and power demand}
  \label{pic:ACIVP} 
  \vspace{-.5cm}
\end{figure} 
Fig. \ref{pic:ACIVP} represents this process for (\ref{eq:PEMOpMo}). Similarly, we have an equivalent process that applies to (\ref{eq:AELOpMo}). So, we can update (\ref{eq_Water_2}) - (\ref{Convwatp3}), and (\ref{eq_H2_6}) and replace binary variables by their optimal values. Now, the surrogate optimization problem is much easier than the original one because it has a smaller number of constraints and contains only continuous variables. We define the optimal strategy as a set of binary variables in the optimal values and active constraints. ACIVP maps the input data, including real-time and forecasted values of wind speeds, water demand, and power demand, to the corresponding optimal strategy. The relationship between input data and output data is explored by using a ML technique such as the decision tree, which has good performance in classification: 
\begin{equation} \label{eq:superMLF}
\psi = \mathcal{H}(\varphi),
\end{equation}
where $\varphi$ and $\psi$ are the input and output of the prediction model, respectively. The prediction model uses $\Gamma =\{(\varphi_1,\psi_1),(\varphi_2,\psi_2),...,(\varphi_{\mathcal{N}_\mathrm{d}},\psi_{\mathcal{N}_\mathrm{d}})\}$ as a dataset to do the training task. The training dataset, which is created by solving offline optimization problems, is required to develop the hypothesis function. An accurate hypothesis function directly depends on the quality of the training dataset, so it is vital to build a reliable, robust dataset. As a result of solving $\mathcal{N}_\mathrm{d}$ offline optimization problems, we obtain $\mathcal{N}_\mathrm{d}$ independent samples and create a training dataset. This dataset includes optimal strategies containing optimal binary variable values and active constraints. The sampling process continues until there is a large enough set of distinct strategies. We use the missing strategies bound method for the sampling procedure to build a training data set that covers all probable strategies \cite{bertsimas2021voice}.
Once we have completed sampling and creating the dataset, we map its input data to its output data using the prediction function in (\ref{eq:superMLF}). Algorithm \ref{alg:OCTsOMIO} shows the proposed fast solution method based on ACIVP for the real-time optimal operation of W2H-LCCI. The input data includes the set of the wind speed, water demand, and power demand and $\psi= \{\textbf{b}_t^\textrm{\textbf{p*}}, \textbf{b}_t^\textrm{\textbf{we*}},\textbf{b}_t^\textrm{\textbf{des*}},\textbf{b}_t^\textrm{\textbf{fc*}},\gamma_t^\textrm{S}\}$ represents the related strategy.
\begin{algorithm}[!h]
\label{alg:OCTsOMIO}
  \caption{\footnotesize ACIVP method for the real-time optimal operation}
  \footnotesize 
\begin{algorithmic}[1]
\justifying
\STATE Input data: wind speed, water demand, and power loads ($\varphi$)
\STATE Use (\ref{eq:superMLF}) to find the optimal values for binary variables and \\ active constraints $\psi = \{\textbf{b}_t^\textrm{\textbf{p*}}, \textbf{b}_t^\textrm{\textbf{we*}},\textbf{b}_t^\textrm{\textbf{des*}},\textbf{b}_t^\textrm{\textbf{fc*}},\gamma^\textrm{S}\}$.
\STATE Set the integer variables to the optimal values that are found in \\\textbf{Step 2} and update (\ref{eq_Water_2}) - (\ref{Convwatp3}), and (\ref{eq_H2_6})
\STATE Remove the redundant constraints based on \textbf{Step 2}
\STATE Develop a new optimization problem based on \textbf{Step 3} and \textbf{Step 4}
\STATE Solve the optimization problem obtained in \textbf{Step 5}, which is smaller \\ and continues, to obtain the optimal operation of the W2H-LCCI.
\end{algorithmic}
\end{algorithm}

\subsection{High-fidelity Method by Enhancing ACIVP with FSE}

The discussed ACIVP method was originally proposed in the domain of optimization theory \cite{bertsimas2021voice,bertsimas2022online}, however to our knowledge, it has not been applied to the fields of sustainable energy. Although the ACIVP-based method solves MCIPs very quickly, it relies on its ability to accurately prediction of active constraints and optimal values of the binary variables. Predicting active constraints and binary variables is much more challenging than predicting something that follows clear patterns, like power demands, since a relationship between inputs, including wind speed, water demand, and power demand, and outputs, including active constraints and binary variables, are not yet well established. There can be no trust in optimization results if accuracy is not high enough, regardless of how fast they are generated. The ACIVP accuracy may not be sufficiently high in large-scale optimization problems \cite{bertsimas2022online}. Since (\ref{eq:AELOpMo}) and (\ref{eq:PEMOpMo}) are large-scale optimization problems, improving prediction accuracy is essential.

\begin{figure}[!t]  
  \centering
\includegraphics[width=0.489\textwidth]{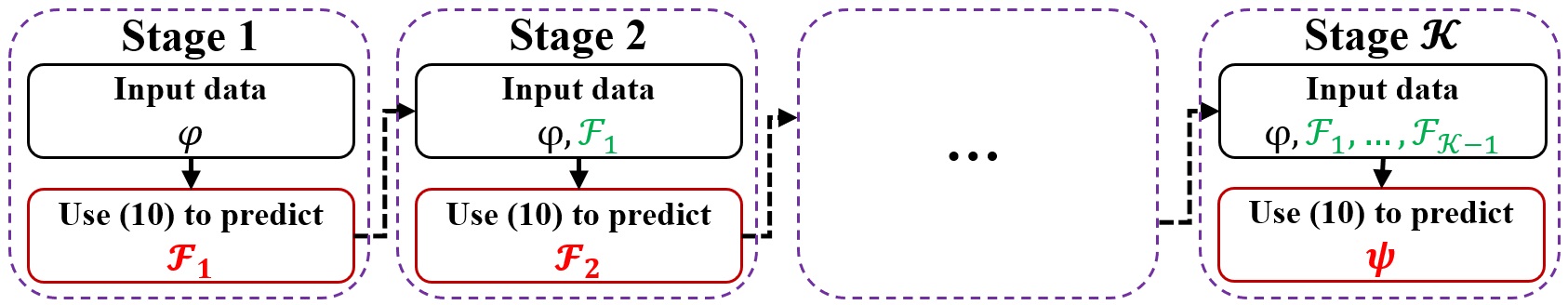}
 \centering
     \vspace{-.5cm}
    \caption{ACIVP-FSE procedure for improving the prediction accuracy}
  \label{pic:ACIVPFSE}    
       \vspace{-.5cm}
\end{figure}

To improve ACIVP accuracy and make its outcomes more reliable, we propose a multi-stage ACIVP-FSE method, which is also a significant contribution to this paper. The addition of more features can enhance learning performance. Therefore, by expanding the feature space, we can improve accuracy \cite{liu2019feature}. In the proposed multi-stage ACIVP-FSE method, the input and output layers of each stage are designed with different learning targets, namely newly developed features, to correct mapping model learning errors. Through solving offline optimization problems, various types of raw data are generated. An additional feature is introduced at each stage of the ACIVP-FSE using raw data to improve learning accuracy. It depends on the learning targets to determine how many stages are needed in each case. Similar to (\ref{eq:superMLF}), the prediction
functions map the input data of each stage to the output data. These prediction functions for ACIVP-FSE are shown in (\ref{eq:predictfuncFSE}). 
\begin{equation}
\label{eq:predictfuncFSE}
\begin{cases}
\mathcal{F}_1 = \mathcal{H}^\textrm{FSE}_1(\varphi), & \textrm{Stage 1} \\
\mathcal{F}_2 = \mathcal{H}^\textrm{FSE}_2(\{\varphi,\mathcal{F}_1\}),  & \textrm{Stage 2} \\
\vdots \\
\psi = \mathcal{H}^\textrm{FSE}_\mathcal{K}(\{\varphi,\mathcal{F}_1,\mathcal{F}_2,\ldots,\mathcal{F}_{\mathcal{K}-1}\}),  & \textrm{Stage } \mathcal{K}
\end{cases} 
\end{equation} 
$\mathcal{F}_1$ is predicted as the first stage output. $\mathcal{F}_1$ is then added as an additional feature in stage two to enhance the accuracy of $\mathcal{F}_2$, which is an output of the second stage. At the end of the final stage, all of the output data ($\psi$) are predicted. Algorithm \ref{alg:ACIVP-FSE} and Fig. \ref{pic:ACIVPFSE} show the multi-stage ACIVP-FSE method for improving the prediction accuracy of the optimal strategy. 
\begin{algorithm}[!h]
\label{alg:ACIVP-FSE}
  \caption{\footnotesize ACIVP-FSE procedure for improving \\ the prediction accuracy of the optimal strategy ($\psi$)}
  \footnotesize 
\begin{algorithmic}[1]
\justifying
\STATE Input data: wind speed, water demand, and power loads ($\varphi$)
\STATE Define a new feature ($\mathcal{F}_1$) and predict it base on (\ref{eq:predictfuncFSE}): \\ $\mathcal{F}_1 = \mathcal{H}^\textrm{FSE}_1(\varphi)$
\FOR{i = 1:${\mathcal{K}-2}$ }
\STATE Develop new features using (\ref{eq:predictfuncFSE}):  $\mathcal{F}_{i+1} = \mathcal{H}^\textrm{FSE}_{i+1}(\varphi,\mathcal{F}_{1},\ldots,\mathcal{F}_{i})$
\ENDFOR
\STATE Predict the optimal strategy by using all of the developed features: \\ $\psi = \mathcal{H}^\textrm{FSE}_\mathcal{K}(\{\varphi,\mathcal{F}_1,\mathcal{F}_2,\ldots,\mathcal{F}_{\mathcal{K}-1}\})$
\end{algorithmic}
\end{algorithm}

ACIVP uses (\ref{eq:superMLF}) to predict the optimal values of all binary variables and the set of active constraints in one stage. While ACIVP-FSE includes $\mathcal{K}$ different stages and uses (\ref{eq:predictfuncFSE}) to improve the accuracy of the prediction. The multi-stage ACIVP-FSE method consists of $\mathcal{K}$ stages to expand the feature space. In the first stage, the proposed method predicts a portion of binary variables in $\mathcal{F}_1$, which is used as an additional feature in the next stage. The input data of stage 2 will be updated with this value to improve its accuracy in the prediction of $\mathcal{F}_2$ which contains another portion of binary variables. The process will continue to the last stage, where all binary variables and active constraints will be predicted more accurately. After that, the related constraints will be updated and the surrogate problem will be developed. To conduct a successful analysis, it will be necessary to revise the initial dataset into $\mathcal{K}$ datasets using the following approach: 
$\Gamma^\textrm{FSE} =$
\begin{equation}
\label{eq:datasetmulti}
 \\
\begin{cases}
\resizebox{.5\hsize}{!}{$(\{\varphi_1\},\{\mathcal{F}_{1,1}\}), \ldots ,(\{\varphi_{\mathcal{N}_d}\},\{\mathcal{F}_{1,{\mathcal{N}_d}}\})$} & \textrm{\small Stage 1} \\
\resizebox{.65\hsize}{!}{$(\{\varphi_1,\mathcal{F}_{1,1}\},\{\mathcal{F}_{2,1}\}),\ldots, (\{\varphi_{\mathcal{N}_d},\mathcal{F}_{1,\mathcal{N}_d}\},\{\mathcal{F}_{2,\mathcal{N}_d}\}),$}  & \textrm{\small Stage 2} \\
\vdots \\
\resizebox{.8\hsize}{!}{$(\{\varphi_1, \ldots,\mathcal{F}_{\mathcal{K}-1,1}\}, \{\mathcal{F}_{\mathcal{K},1}\}), \ldots, (\{\varphi_{\mathcal{N}_d}, \ldots,\mathcal{F}_{\mathcal{K}-1,\mathcal{N}_d}\}, \{\mathcal{F}_{\mathcal{K},\mathcal{N}_d}\}),$}  & \textrm{\small Stage } \small \mathcal{K}
\end{cases} \nonumber
\end{equation}

This paragraph will discuss why multi-stage ACIVP-FSE can improve accuracy. Generating more features can boost learning performance and improve accuracy \cite{liu2019feature}. It means that the more diverse and relevant features we have, the more robust and effective the learning process will be. Therefore, it is critical to carefully consider the features used in the learning algorithm and explore ways to augment them for optimal results.
An additional feature is introduced at each stage of the ACIVP-FSE using raw data to improve learning accuracy. Despite some learning errors in the ACIVP mapping model, these errors reduce gradually in the multi-stage ACIVP-FSE mapping model. The proposed architecture aims to incorporate more features and expand feature space to improve learning accuracy. The multi-stage ACIVP-FSE method may have different stages and features depending on the application, as explained in the following subsection.



\subsection{Stage Design for Multi-Stage ACIVP-FSE}
This section proposes different methods for designing the multi-stage ACIVP-FSE to replace the ACIVP approach. After solving offline optimization problems, a considerable amount of raw data becomes available for developing new features. New features can be created by incorporating raw data related to binary variables and active constraints. Different scenarios are proposed to create new features, determine the number of stages and develop the specific multi-stage ACIVP-FSE. 
\subsubsection{Physical Attributes-Based Design}
New features can be created through consideration of the physical interpretation of the problem. The suggested design is beneficial for solving a problem that involves various types of equipment with binary variables in their models. We start by predicting binary variables related to a specific piece of equipment. This prediction is then used as a feature to enhance the accuracy of predicting another portion of binary variables associated with other types of equipment. This procedure will continue until the last stage.
For example, the mathematical model of W2H-LCCI contains binary variables for four equipment types: pumps, desalination, fuel cells, and electrolysis. Therefore, We divide the binary variables into four groups based on their physical meaning, and each group is assigned to a specific stage. The five new features are $\{\mathcal{F}_1, \mathcal{F}_2, \mathcal{F}_3, \mathcal{F}_4, \mathcal{F}_5\}  = \{\textbf{b}_{t}^{\textbf{p*}},\textbf{b}_{t}^{\textbf{des*}},\textbf{b}_{t}^{\textbf{fc*}},\textbf{b}_{t}^{\textbf{we*}},\gamma_t^\textrm{S}\}$. We develop a five-stage ACIVP-FSE for our optimization model, which improves the accuracy of the ACIVP method.
After the prediction of binary variables and active constraints, we update (\ref{eq_Water_2})-(\ref{Convwatp3}) and (\ref{eq_H2_6}), and remove (\ref{eq_H2_8}) and (\ref{eq:AEL}). We can now replace optimization problems (\ref{eq:AELOpMo}) and (\ref{eq:PEMOpMo}) with their surrogate problems containing only continuous variables and active constraints.

\subsubsection{Time-Based Design}
Another approach to dividing binary variables and active constraints is based on time. A dataset consisting of $\mathcal{T}$ time steps is examined, and binary variables are divided into $\mathcal{T}$ distinct groups, thereby generating $\mathcal{T}$ novel features. This design can be suitable for solving problems that involve multi-periodic operations. The binary variables associated with the first time step will be predicted and used as supplementary features in the second stage. This process will persist until the final stage.
 Upon completing the prediction task, we update the relevant constraints and develop a surrogate optimization problem. 
For example, the W2H-LCCI model has four types of binary variables over $\mathcal{T}$ time steps. We group the binary variables into $\mathcal{T}$ groups based on time and assign each group to a specific stage. Therefore, the set of new features is $\mathcal{F}_k = \{\textbf{b}_{k}^{\textbf{p*}},\textbf{b}_{k}^{\textbf{des*}},\textbf{b}_{k}^{\textbf{fc*}},\textbf{b}_{k}^{\textbf{we*}},\gamma_k^\textrm{S}\}, \forall \,\, k \in \{1,2,..,\mathcal{T}\}$.
Thus, we propose a $\mathcal{T}$-stage ACIVP-FSE to enhance the precision of the ACIVP method. Following the forecasting of binary variables and active constraints, we update the related constraints and develop the surrogate problem.

\subsubsection{Number of Binary Variables-Based Design}
Another method for categorizing binary variables involves dividing them based on their quantity. Our proposal suggests dividing the binary variables into $\mathcal{K}$ groups, where each group has an equal number of binary variables, regardless of their physical interpretation. This process results in $\mathcal{K}$ separate stages within the dataset, each containing $\mathcal{K}$ distinct features.
Using the $\mathcal{K}$-stage dataset, we create a surrogate optimization problem by incorporating (\ref{eq:predictfuncFSE}).
Since this design is based on the number of facilities with binary variables in their models, it will differ for different W2H-LCCI case studies.

\subsubsection{Coupled-Based Design}
We propose to predict a certain binary variable and use it as a new feature for the prediction of another binary variable that shares some high degree of interdependence. The new features could provide a more comprehensive understanding of the underlying relationships and improve accuracy and performance. For example,  (\ref{eq_H2_8}) indicates a strong relationship between binary variables of FC units and water electrolysis in a W2H-LCCI operation. Predicting one variable early and using it as an additional feature for predicting the other can improve prediction accuracy. Specifically, we use $\mathcal{F}_1 = \textbf{b}^{\textbf{fc*}}$ to enhance $\textbf{b}^{\textbf{we*}}$ prediction accuracy. A similar approach is employed for predicting $\textbf{b}^{\textbf{p*}}$ and $\textbf{b}^{\textbf{des*}}$. After prediction, we develop a surrogate optimization problem using (\ref{eq:predictfuncFSE}).

\subsection{Stage Ordering Strategy for Multi-Stage ACIVP-FSE}
We employ the multi-stage ACIVP-FSE method to enhance ACIVP accuracy, which depends on the stage order design. If FSE is first applied to a feature with low prediction accuracy, ACIVP-FSE accuracy may be even lower than ACIVP accuracy. Consequently, to achieve optimal performance with ACIVP-FSE, the stage order must be carefully considered. Our proposal involves utilizing data from the ACIVP method to identify the most suitable features for initiating the ACIVP-FSE process. We used the results of the ACIVP method to determine the accuracy prediction of various features. The feature with the highest accuracy will be selected as $\mathcal{F}_1$. This feature has minimal negative impact on the input data of the second stage, and can potentially enhance the accuracy of $\mathcal{F}_2$. We arrange the features in order of their accuracy. We then assign the feature with the highest predicted accuracy to the first stage of the multi-stage ACIVP-FSE. This process will continue to the last stage to improve prediction accuracy. Some stage designs, such as coupled-based or time-based, require only finding $\mathcal{F}_1$ since the features of the upcoming stage are related to the previous stages.

\section{Case Study} \label{sec:casestudy}
\subsection{Introducing Test beds}
\begin{figure}[!b]  
    \vspace{-.6cm}
  \centering
\includegraphics[width=0.4\textwidth]{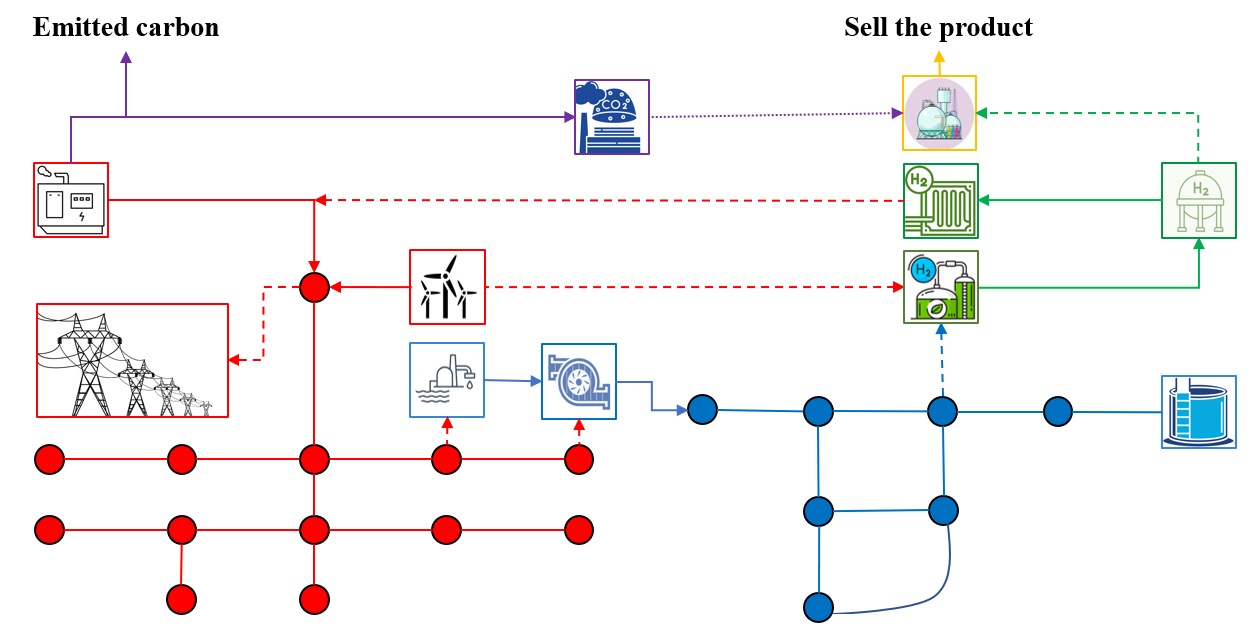}
 \centering
     \vspace{-.2cm}
    \caption{W2H-LCCIs for a small stand-alone coastal city}
  \label{pic:LCWENSACoast}    
\end{figure}

\begin{figure}[!t]  
  \centering
\includegraphics[width=0.489\textwidth]{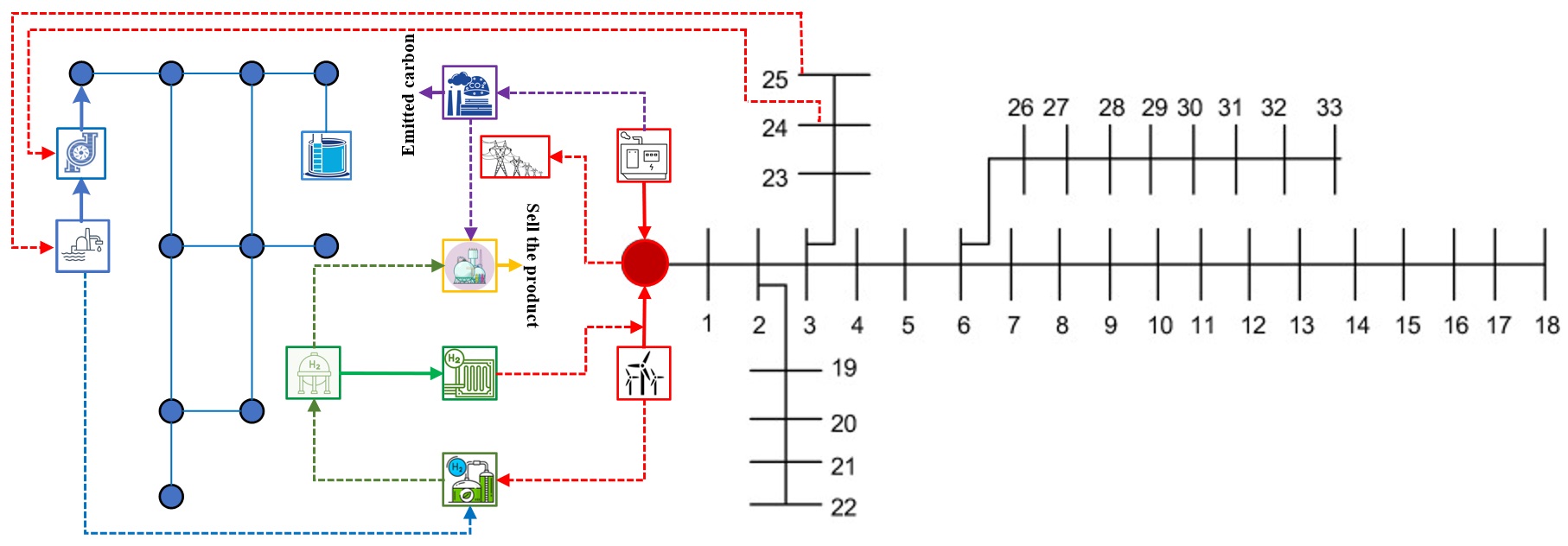}
 \centering
 \vspace{-.6cm}
    \caption{W2H-LCCI system for a small stand-alone island}
  \label{pic:LCWENSACoastIEEE33}   
      \vspace{-.7cm}
\end{figure}

We demonstrate the robustness of the proposed method by examining it on two test beds as shown in Fig. \ref{pic:LCWENSACoast} and Fig. \ref{pic:LCWENSACoastIEEE33}. 
A modified IEEE 13-bus system with an 8-node EPANET WDS which can represent a small stand-alone coastal city, and the IEEE 33-bus system with a 13-node Otsfeld WDS which can be used for a remote small island. CIs are distinct entities on a larger scale, and they have conflicts between them, making their collaboration impractical. Training data sets are built by solving offline optimization problems using wind speed and load data. 24-hour wind speed curves are extracted between 2008 and 2022 \cite{meteoblue} and interpolated using spline interpolation every five minutes. 24-hour nodal load curves of power demand, excluding water pumps and desalination, are also extracted \cite{dataminer2} and interpolated every five minutes. Upon dataset construction, we evaluate the ACIVP method's resilience using real data, including actual wind speed and load data, derived from Midcontinent Independent System Operator (MISO) \cite{Misowind}. For each case study, we utilize these data to determine the optimal operation of W2H-LCCI, incorporating two distinct water electrolysis technologies. For comparison, the conventional method solution time is compared to the ACIVP and ACIVP-FSE methods. We also compare the accuracy of the ACIVP-FSE method with the ACIVP method based on different stage designs.

\subsection{Importance of Online Optimal Operation for W2H-LCCI}
\begin{figure}[!b] 
  \vspace{-.5cm}
    \includegraphics[width=0.225\textwidth]{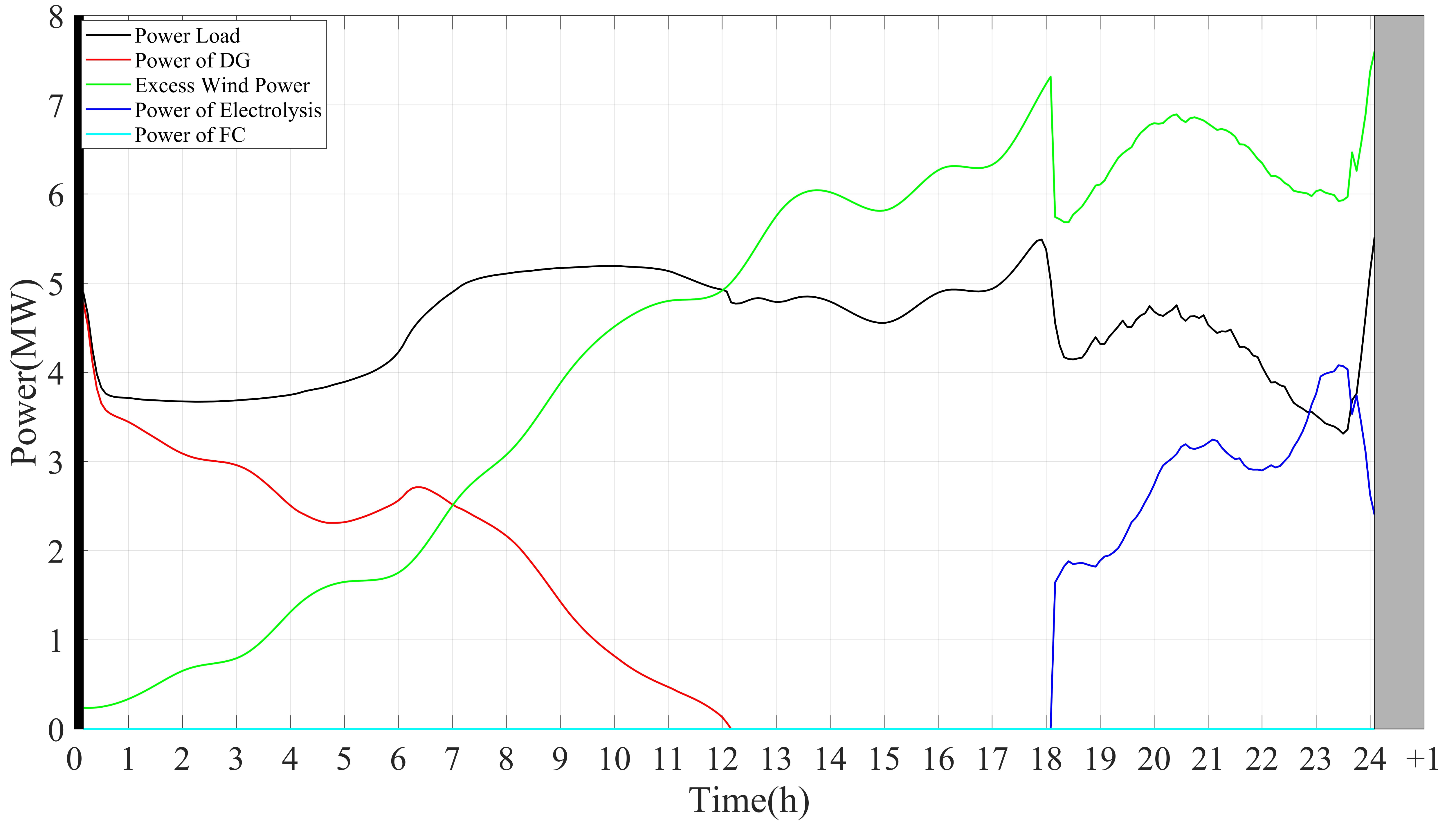} \,
    \includegraphics[width=0.225\textwidth]{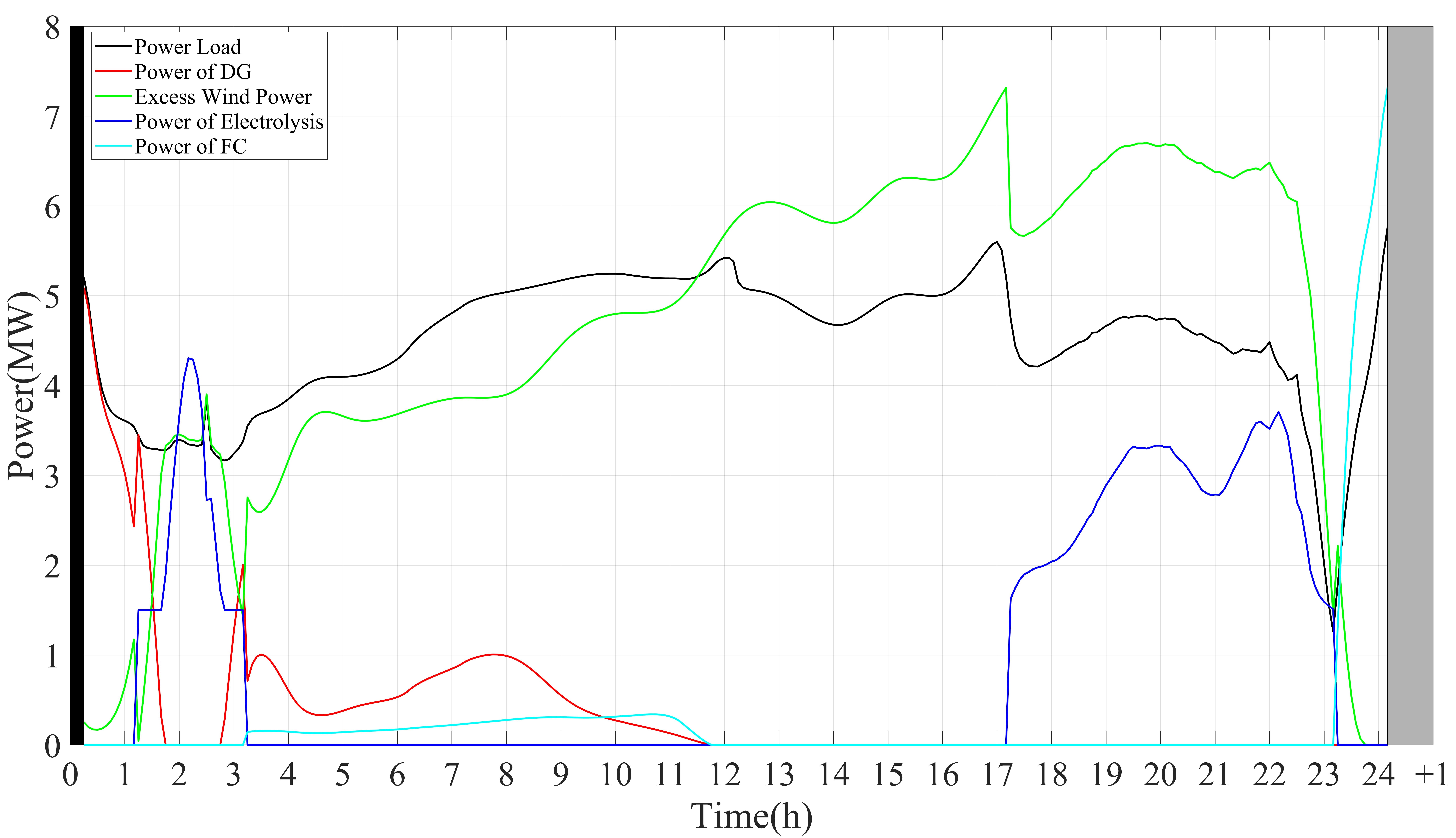}\\
    \includegraphics[width=0.225\textwidth]{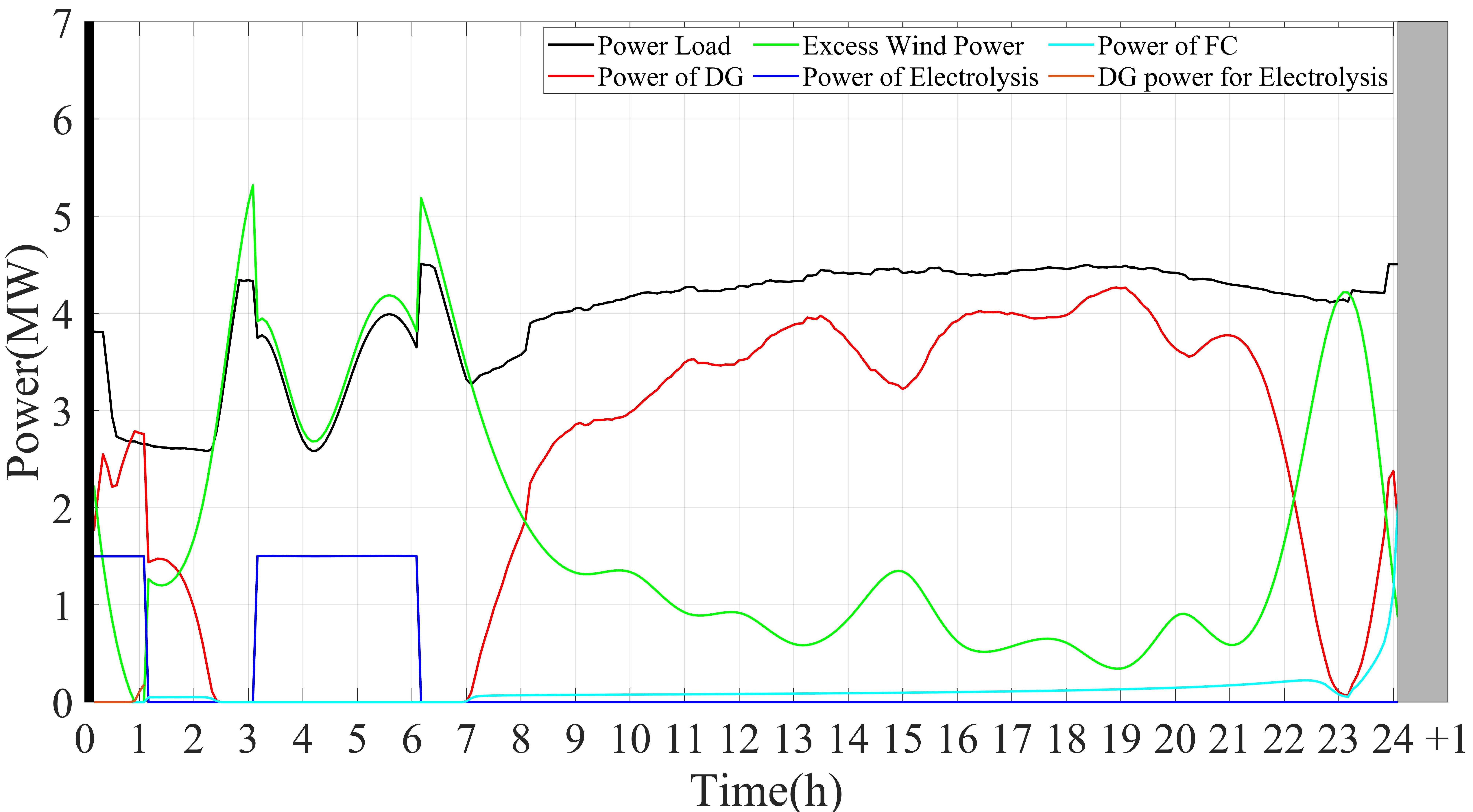} \,
    \includegraphics[width=0.2273\textwidth]{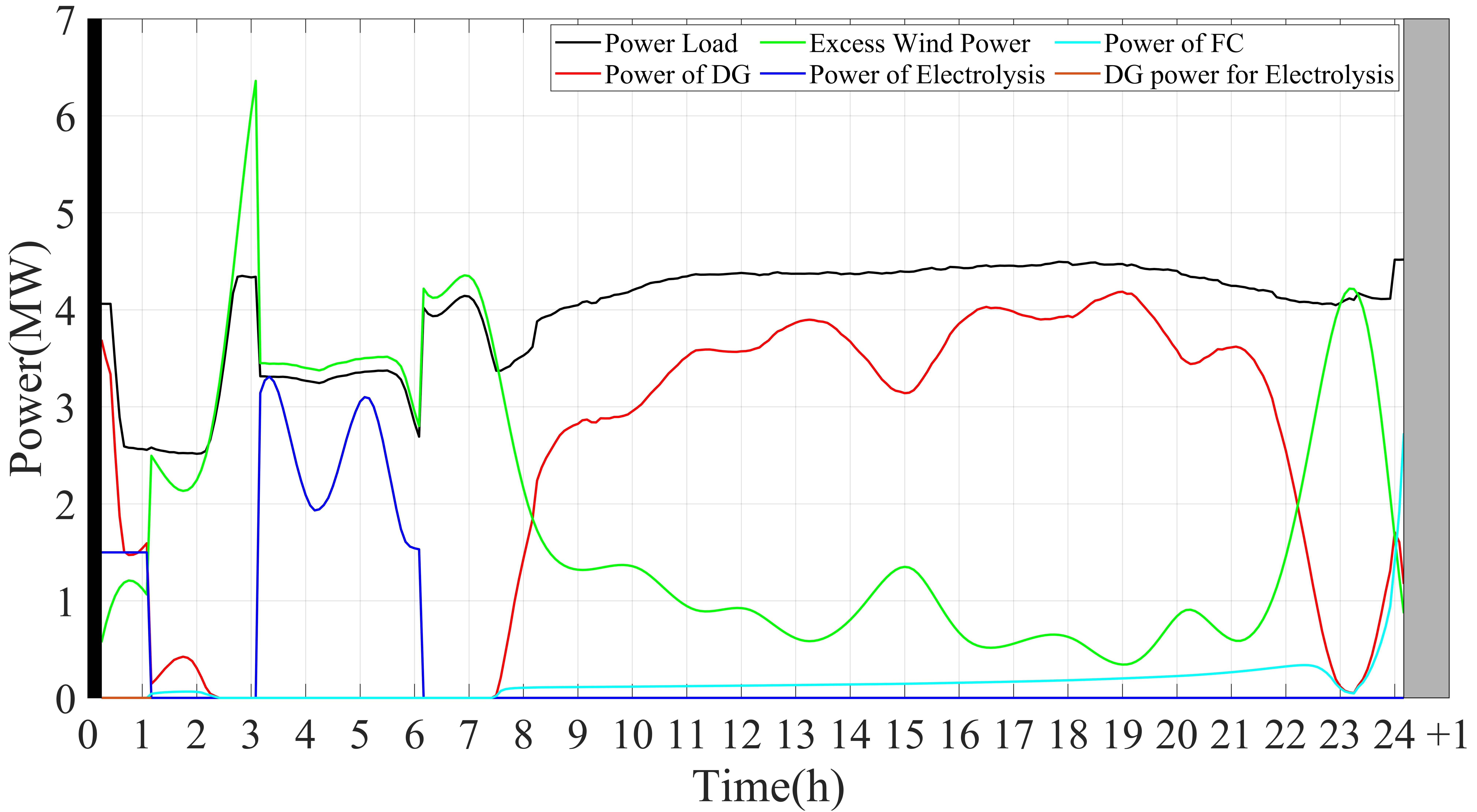}\\
    \vspace{-.4cm}
  \caption{First W2H-LCCI in two consecutive time steps (00:10, 00:15) and two case studies: First (top row), Second (bottom row)}
   \label{pic:casestudy}
\end{figure}

We apply the real-time optimization models introduced in section \ref{sec: LCEWH_Optmodel} to the proposed W2H-LCCIs (illustrated in Fig. \ref{pic:LCWENSACoast} and Fig. \ref{pic:LCWENSACoastIEEE33}) to obtain optimal W2h-LCCI operation. According to the optimal operation of W2H-LCCI for the two consecutive time intervals, Fig. \ref{pic:casestudy} demonstrates the results.
In the presented system, the power load is denoted by the color black, including various components such as the water pump, water desalination, and other power loads that are connected to the PDN. The diesel power generation is represented by red color. The available wind energy after fulfilling the water electrolysis demand is shown by the green color. The blue color represents the power demand for water electrolysis, while the cyan color indicates the power of the FC units. Besides, the orange color represents the power of the diesel generator used for AEL water electrolysis when there is insufficient wind energy. Adjusting the wind speed prediction in the second time interval alters the optimal operation of water electrolysis. Fig. \ref{pic:casestudy} highlights the necessity for real-time optimal operation of W2H-LCCI, as any variance between the forecasted and actual wind speeds can significantly alter the optimal operation.

\subsection{Solution Time Evaluation of ACIVP and ACIVP-FSE}
Since any discrepancy between predicted and actual wind speeds significantly influences the optimal operation of W2H-LCCI, it is critical to promptly solve the optimization problem to obtain real-time optimal operation. Solving the related optimization problems through conventional approaches is a lengthy process that requires over 15 minutes. Employing ACIVP, we can replace the initial problem with a surrogate one, and achieve optimal real-time operation of W2H-LCCI in just a few seconds. Table \ref{tab:soltimefirstcase} compares the solution times of conventional approaches and the ACIVP method for two case studies. This table illustrates the superior efficacy of the ACIVP method over existing methods. ACIVP-FSE solves the optimization problem in a similar timeframe to ACIVP. The only difference between these methods is their accuracy in finding optimal values for binary variables and active constraints, which we will compare in Section \ref{sec:Accuracyevaluation}.
\begin{table}[!h]
      \vspace{-.4cm}
\centering
\caption{Solution Time of Optimization Methods}
\vspace{-.2cm}
\label{tab:soltimefirstcase}
\footnotesize
\begin{tabular}{ccccc}
\hline\hline
\multirow{2}{*}{Case Study}        &\multirow{2}{*}{Time}     & \multicolumn{3}{c}{Solution Time (sec)}                    \\ \cline{3-5} 
             &                & \multicolumn{1}{c}{Conventional Method} & ACIVP & ACIVP-FSE \\ \hline \hline
\multirow{2}{*}{First} &00:10 & \multicolumn{1}{c}{892.16}              & 1.82        & 1.63   \\ \cline{2-5}
&00:15  & \multicolumn{1}{c}{914.53}                   & 1.32       &1.83      \\ 
\hline \hline
\multirow{2}{*}{Second}  &00:10 & \multicolumn{1}{c}{1263.74}              & 4.09       &3.67 \\ \cline{2-5}
&00:15  & \multicolumn{1}{c}{1986.70}                   & 2.96      &3.11     \\ 
\hline\hline
\end{tabular}
\vspace{-.7cm}
\end{table}

\subsection{Accuracy Evaluation of ACIVP and ACIVP-FSE} \label{sec:Accuracyevaluation}
This section evaluates various stage designs proposed to create a multi-stage ACIVP-FSE. The first is a physical attributes-based design, resulting in a five-stage ACIVP-FSE with stages related to the pump, water desalination, FC units, water electrolysis, and active constraints. The second design is based on time, dividing binary variables into 288 time intervals of 24 hours. This results in a 289-stage ACIVP-FSE with 288 stages related to binary variables and one stage related to active constraints. The third design is based on the number of binary variables. Binary variables are divided into two groups to create a three-stage ACIVP-FSE. The last one is a coupled-based design, combining FC units and water electrolysis into one group and water pumps and desalination into another, resulting in a three-stage ACIVP-FSE with two stages related to binary variables and one stage related to active constraints. Table \ref{tab:accuracy} displays the accuracy of each design. This Table shows that ACIVP-FSE increases the accuracy of ACIVP in the prediction of binary variables and active constraints. Out of four different design types that were suggested, the physical attributes-based design shows the greatest level of precision with a score of $98.87\%$
 \begin{table}[!h]
      \vspace{-.4cm}
\centering
\caption{Evaluation of ACIVP and ACIVP-FSE: Accuracy Analysis}
\vspace{-.2cm}
\label{tab:accuracy}
\footnotesize
\begin{tabular}{cccccc}
\hline\hline
\multirow{2}{*}{Case Study} & \multirow{2}{*}{ACIVP($\%$)} & \multicolumn{4}{c}{ACIVP-FSE($\%$)}                                                                      \\ \cline{3-6} 
                            &                        & \multicolumn{1}{c}{Physical} & \multicolumn{1}{c}{Time}  & \multicolumn{1}{c}{Number} & Coupled \\ \hline\hline
I                           & 77.58                  & \multicolumn{1}{c}{98.87}    & \multicolumn{1}{c}{93.44} & \multicolumn{1}{c}{86.63}  & 94.17    \\ \hline
II & 74.91 & \multicolumn{1}{c}{97.46} & \multicolumn{1}{c}{92.17} & \multicolumn{1}{c}{85.71} & 93.46 \\ \hline
\end{tabular}
\vspace{-.4cm}
\end{table}

\section{Conclusion} \label{sec: conclusions}
This paper proposes real-time optimization models for W2H-LCCI consisting of power, water, and hydrogen systems to reduce carbon emissions while maximizing wind energy utilization. W2H-LCCI utilizes wind power to generate hydrogen through electrolysis and combines it with carbon capture to reduce carbon emissions from the power sector. To achieve optimal operation, the paper develops convex mathematical models and control strategies dependent on water electrolysis technology, which are MICPs. It is computationally challenging to solve the MCIPs. A novel mapping-based approach called AVIVP is proposed to solve this problem quickly to address wind energy and power demand intermittent nature. ACIVP predicts the binary variable values and the set of limited-number constraints, which most likely contain all of the active constraints, based on historical optimization data and surrogates the initial MICP with a small-scale continuous convex optimization problem that can be solved quickly. Besides, we improve the accuracy of the ACIVP method by applying FSE and developing a multi-stage ACIVP-FSE. Four different stage designs of multi-stage ACIVP-FSE and stage ordering are proposed and compared to improve the ACIVP method. The proposed system and solution method are validated through case studies on the IEEE 13-bus power system with the EPANET 8-node water system and the IEEE 33-bus power system with the Otsfeld 13-node water system. The results demonstrate that the ACIVP approach significantly reduces solution time, by a few seconds. This allows for real-time optimal W2H-LCCI operation and overcomes intermittent wind energy and power demand. The ACIVP-FSE method can result in a $27.44\%$ improvement in ACIVP accuracy.

\bibliographystyle{IEEEtran}	
\bibliography{References}
\end{document}